\documentclass[aps, prb, amssymb, reprint,superscriptaddress]{revtex4-1}
\usepackage{amsmath}
\usepackage{amsfonts}
\usepackage{amssymb}
\usepackage{url}
\usepackage[colorlinks=true,linkcolor=blue,citecolor=blue]{hyperref}
\usepackage{color}
\usepackage{epsfig}
\usepackage{subfigure}
\usepackage{exscale}
\usepackage{braket}
\usepackage{xcolor}

\usepackage{float}
\usepackage{verbatim}

\begin{document}

\title{Quantum Simulation and Spectroscopy of Entanglement Hamiltonians}
\author{M. Dalmonte}
 \affiliation{International Center for Theoretical Physics, 34151 Trieste, Italy}
\author{B. Vermersch}
    \affiliation{Institute for Theoretical Physics, University of Innsbruck, A-6020 Innsbruck, Austria}
    \affiliation{IQOQI of the Austrian Academy of Sciences, A-6020 Innsbruck, Austria}
\author{P. Zoller}
    \affiliation{Institute for Theoretical Physics, University of Innsbruck, A-6020 Innsbruck, Austria}
    \affiliation{IQOQI of the Austrian Academy of Sciences, A-6020 Innsbruck, Austria}

\begin{abstract}
Entanglement is central to our understanding of many-body quantum matter. In particular, the entanglement spectrum, as eigenvalues of the reduced density matrix of a subsystem, provides a unique footprint of properties of strongly correlated quantum matter from detection of topological order to characterisation of quantum critical systems. However, direct experimental measurement of the entanglement spectrum has so far remained elusive due to lack of direct experimental probes. Here we show that the entanglement spectrum of the ground state of a broad class of Hamiltonians becomes directly accessible as quantum simulation and spectroscopy of an entanglement Hamiltonian, building on the Bisognano-Wichmann (BW) theorem of axiomatic quantum field theory. Remarkably, this theorem gives an explicit physical construction of the entanglement Hamiltonian, identified as Hamiltonian of the many-body system of interest with spatially varying couplings. Building on this, we propose an immediate, scalable recipe for implementation of the entanglement Hamiltonian, and measurement of the corresponding entanglement spectrum as spectroscopy of the Bisognano-Wichmann Hamiltonian with synthetic quantum systems, including atoms in optical lattices and trapped ions. We illustrate and benchmark this scenario on a variety of models, spanning phenomena as diverse as conformal field theories, topological order, and quantum phase transitions.
\end{abstract}

\maketitle

\begin{figure}
\centering \includegraphics[width=0.93\columnwidth]{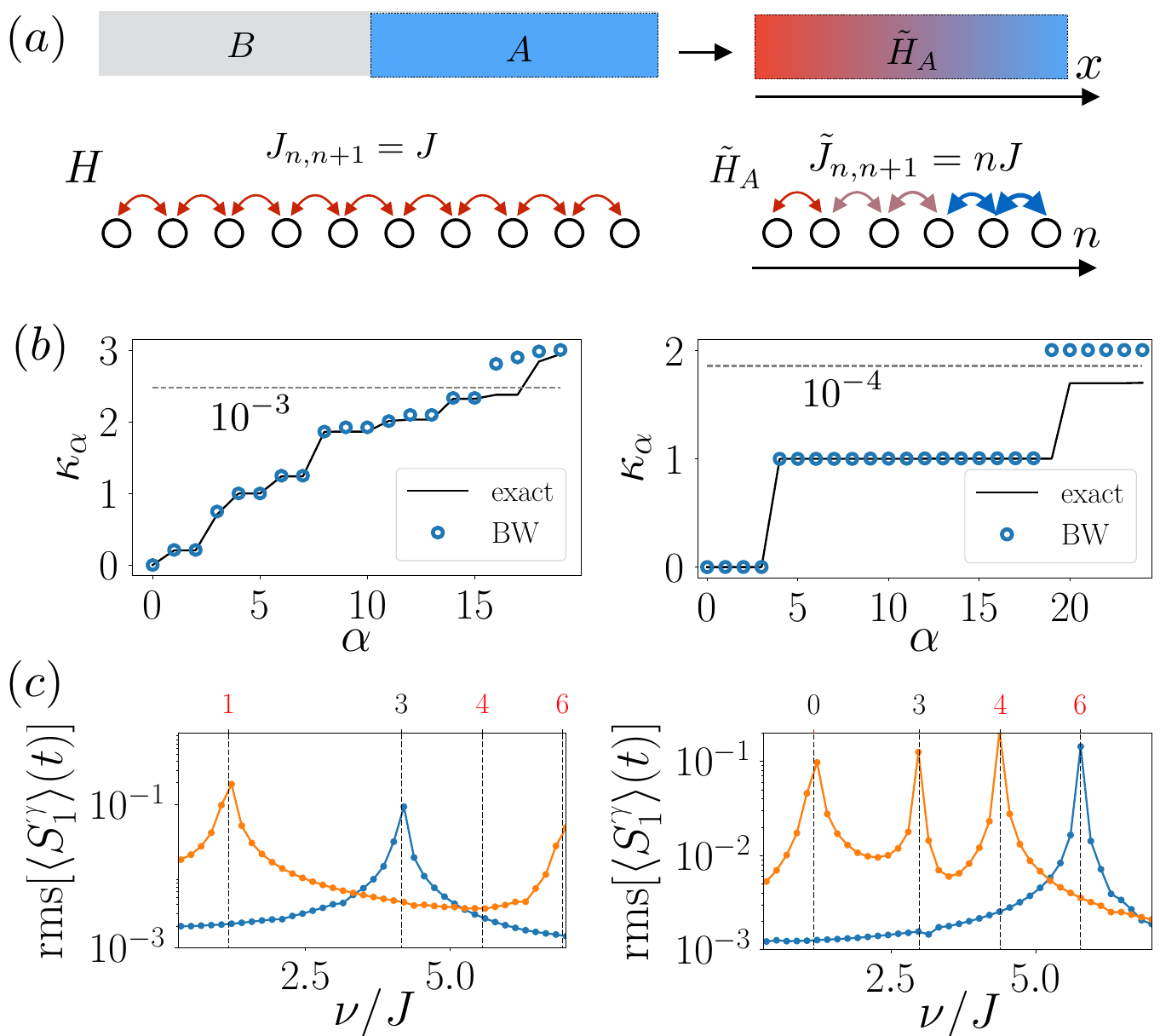}
\caption{{\it Entanglement spectra via spectroscopy.} 
(a) We are interested in the entanglement properties of the ground state of a given Hamiltonian $H$ and bipartition A. The corresponding entanglement Hamiltonian is given by Eq.~\eqref{eqBW}, which on a lattice can be recast as couplings with increasing magnitude as a function of the distance from the boundary.
(b) Illustration of the accuracy of the lattice Bisognano-Wichmann (BW) prediction for the Spin $1$ XXZ chain in the Haldane phase. The dimensionless ratios $\kappa_\alpha$ are represented as a function of the eigenvalue index $\alpha$, with $\alpha_0=4$.
Dashed lines represent the level of reference for the corresponding eigenvalues $\tilde\lambda$.
Left panel: For a system size $L'=8$, $L=100$, and $\Delta=0.3$, the prediction of BW is excellent despite the small size of the bipartition.
Right panel: For $\Delta=1$, $L'=40$ and $L=80$ (PBC), the BW perfectly predicts the characteristics degeneracies of the Haldane phase.
(c) Spectroscopy of the EH for the parameters of panel (b, left). This is realised by applying a perturbation $h=a\sin(\nu t)S_1^\gamma$ and measuring the response $\langle S_1^\gamma(t)\rangle$ for $\gamma=z,x$, respectively in blue and orange. In the left (right) panel, the system is initialised in the ground (first excited) state. Here $a=0.02J$ and $Jt_{\mathrm{obs}}=150$. The vertical dashed lines correspond to the exact values, and the top label on each line indicates the eigenvalue index $\alpha$ (in red when the eigenvalue is degenerate).
}\label{fig:recipe} 
\end{figure}

Entanglement describes genuinely quantum, non-local correlations between different parts of a physical system~\cite{Amico2008,Eisert2010}. For a system prepared in a pure quantum state $|\Psi\rangle$, entanglement properties are encoded in the reduced density matrix for a subsystem A, defined by $\rho_A = \text{Tr}_B |\Psi\rangle\langle\Psi|$, which we write as 
\begin{eqnarray}
\rho_A = e^{-\tilde{H}_A} =\sum_\alpha e^{-\tilde{\epsilon}_\alpha}|\varphi_\alpha\rangle\langle\varphi_\alpha|.
 \end{eqnarray}
Here $\{\tilde{\epsilon}_\alpha\}$ is a set of eigenvalues, known as entanglement spectrum (ES)~\cite{Li2008,Peschel2009}, and $\tilde{H}_A$ defines an entanglement (or modular) Hamiltonian (EH)~\cite{haag2012local,Regnault:2015aa,Li2008}. The ES plays a paradigmatic role at the interface of entanglement theory and many-body physics~\cite{fradkinbook}, with applications encompassing the characterisation of topological order~\cite{Li2008,Pollmann:2010aa,fidkowski2010entanglement,Regnault:2015aa} to the detection of criticality, quantum phase transitions and spontaneous symmetry breaking~\cite{calabrese2008entanglement,Peschel2009,Cirac:2011aa,Alba:2012aa,Chiara:2012aa,lauchli2013operator}, and the understanding of the efficiency of variational methods based on the tensor network paradigm~\cite{Peschel2009,schollwock2011density}. However, these theoretical insights are at present lacking an experimental counterpart, as the ES has never been experimentally measured due to the lack of probing tools. While the ES could be, at least in principle, measured by full quantum state tomography of $\rho_A$, this is exponentially inefficient with system size. Here, we propose to shift the paradigm of measuring entanglement properties, from {\em a probing of the wave function} (see e.g.~Ref.~\onlinecite{pichler2016measurement,Beverland:2016aa}), to a { \em direct and efficient quantum simulation and spectroscopy of the corresponding EH.} The challenge is, therefore, to develop techniques, accessible in present experiments, which provide a direct realisation of the EH. Below we address this problem, building on the Bisognano-Wichmann (BW) theorem~\cite{bisognano1975duality,bisognano1976duality,guido2011modular} of axiomatic quantum field theory, in the framework of many-body quantum systems.

The BW theorem provides a closed form expression for the EH  $\tilde{H}_A$ for Lorentz invariant quantum field theories. This theorem states that, given a system with Hamiltonian density $H(\vec{x})$ and a half-bipartition, that for simplicity we denote as the subspace with $x_1>0$, the EH  of the ground state of $H$ reads
\begin{equation}\label{eqBW}
\tilde{H}_{A} = 2\pi \int_{\vec{x}\in A} d\vec{x} (x_1 H(\vec{x})) + c',
\end{equation}
with $c'$ a constant to guarantee unit trace of the reduced density matrix. The BW construction holds in any dimensions, and in particular provides a simple explicit form for the EH, which -- as the original physical Hamiltonian -- is built from just local few body terms and interactions. A key feature of this result is that its applicability does not rely on any knowledge of the ground state, and thus can be applied in both gapped and gapless quantum systems, and also at quantum critical points. Moreover, Eq.~\eqref{eqBW} has a clear-cut physical interpretation in terms of {\it entanglement temperature}~\cite{Casini:2011aa}: if we interpret $\rho_A$ as thermal state, this corresponds to a state of the original Hamiltonian $H$ with respect to a locally varying temperature, very large close to the boundary of A, and linearly decreasing far from it. This interpretation has been used, e.g., in the context of Hawking radiation and the Unruh effect ~\cite{SEWELL1982201}. Moreover, the BW theorem has been extended to different geometries~\cite{Casini:2011aa} as well as to real-time dynamics in the presence of additional global symmetries~\cite{Cardy:2016aa}, and may also incorporate gauge symmetries~\cite{Casini:2014aa,Pretko:2016aa}.

While the BW theorem applies strictly speaking only to the ideal scenario of infinite system size and in the continuum, in order to establish connection to condensed matter systems and atomic physics experiments, we will cast it on finite lattice models. Below, we show how this approach is remarkably accurate for many paradigmatic cases in strongly correlated systems, including conformal phases of both spin and fermionic systems, topological phases in one- and two-dimensions, and is able to correctly capture the quantum critical regime of Ising-type models. In all these cases, and in particular, in the ones characterized by infinite correlation lengths (and, as such, potentially more sensitive to finite lattice and finite size effects), we find that the BW entanglement spectrum correctly reproduces the exact low-lying entanglement spectrum as long as Lorentz invariance is approximately realized at low-energies - as a counterexample, we discuss the limitation of this approach for systems with approximately quadratic dispersion relations. 
As a case sample, we illustrate this procedure for the case of spin-1 Heisenberg chains, with Hamiltonian $H_{\textrm{XXZ}}=\sum_{n=-\infty}^{\infty} JH_{n,n+1}$ and Hamiltonian density~\cite{fradkinbook}
\begin{equation}\label{HXXZ}
H_{n,n+1} = S^x_nS^x_{n+1} + S^y_nS^y_{n+1}+\Delta S^z_nS^z_{n+1},
\end{equation}
where $S^\alpha_n$ are spin-1 operators at a site $n$. The lattice BW Hamiltonian (denoted in the following as $\tilde{\mathcal{H}}_A$) is
\begin{equation}
\tilde{\mathcal{H}}_{A,\textrm{XXZ}} = \sum_{n=1}^{\infty}\tilde{J}_{n, n+1} H_{n,n+1}
\end{equation}
with spatially varying coupling strengths (c.f.~Fig.~\ref{fig:recipe}a). Here $\tilde{J}_{n, n+1} = n J$ for a half-system bipartition with open boundary conditions (OBC), and $\tilde{J}_{n, n+1} = J (L'-n)n/L'$ for finite partitions of length $L'$ in the centre of the system with periodic boundary conditions (PBC). In Fig.~\ref{fig:recipe}b, we present typical results for comparison between the exact entanglement spectra for the ground state of $H_{\textrm{XXZ}}$, and the physical spectrum of $\tilde{\mathcal{H}}_{A,\textrm{XXZ}}$. The comparison is drawn by focusing on universal ratios of entanglement eigenenergies, $\kappa_{\alpha; \alpha_0} \equiv  ( \tilde{\epsilon}_\alpha-\tilde{\epsilon}_{0}) / ( \tilde{\epsilon}_{\alpha_0}-\tilde{\epsilon_0} )$, where $\tilde\epsilon_0$ is the lowest entanglement energy in the system (corresponding to the largest eigenvalue of $\rho_A$), and $\tilde \epsilon_{\alpha_0}$ is a reference state (if not explicit, we take the first excited entanglement energy, and define $\kappa_\alpha=\kappa_{\alpha; 1}$). Note that the overall energy scales cancels out in universal ratios. The agreement is excellent even for very modest system sizes for eigenvalues $\tilde\lambda=e^{-\tilde\epsilon_\alpha}$ down to $10^{-4}$, and, deep in the topological phase, the characteristic degeneracy of the ES is captured with errors smaller than $10^{-5}$. Physically building and preparing a synthetic quantum system emulating the BW EH, and performing spectroscopy by probing the system as illustrated in Fig.~\ref{fig:recipe}c will thus provide a direct and efficient measurement of the ES. This approach allows us to exploit and transfer the accuracy and flexibility of conventional spectroscopy to the study of entanglement properties.

\section*{Entanglement Hamiltonians and the Bisognano-Wichmann theorem on a lattice}

The main challenges in applying the BW \eqref{eqBW} theorem to quantum many-body systems in condensed matter physics are the requirements of infinite partitions and Lorentz invariance.
To address the latter, we consider systems on lattices, where Lorentz invariance is often emerging as an effective low-energy symmetry. This correspondence between lattice models and field theory is at the basis of many computational techniques to address continuum problems, such as lattice field theory~\cite{Montvay1994}. The lattice not only provides a natural regularisation, but, for our purposes here, allows us to realise effectively Lorentz invariant dynamics in non-relativistic scenarios such as cold gases. This will come at the price of introducing non-universal effects: as we will show below, those have negligible influence in the universal properties of the ES, and, for sufficiently large systems, they affect only very large eigenenergies.

To be concrete, we recast BW on a finite lattice, and for simplicity consider a 1D lattice of length $2L'$,
\begin{equation}\label{BWL}
H = \sum_{n = -L'}^{L'-1}H_{n,n+1}, \; \tilde{\mathcal{H}}_A = \beta\sum_{n=1}^{L'-1} n H_{n,n+1} + c' + \mathcal{O}_{a, L'}
\end{equation}
with $\beta$ a constant (typically related to the sound velocity of the corresponding low-energy field theory). The last term describes corrections due to finite lattice spacing, and due to the finite size of the sample: these corrections are akin to the ones found when simulating continuum field theories on space-time lattices, as done in various fields including lattice gauge theories~\cite{Montvay1994}. We note that, for the case of the Ising model, the above construction is exact for infinite bipartitions~\cite{Peschel2009},  i.e.~there are no corrections due to the breaking of Lorentz invariance. While we expect that corrections to the universal properties of the ES due to $\mathcal{O}_{a, L'}$ would vanish for sufficiently large systems, experiments are necessarily carried out at finite size. In the following, we address in detail the role of such corrections, by comparing the exact entanglement spectra with the ones obtained via the spectrum of the BW entanglement Hamiltonian for a variety of 1D and 2D models, whose concrete physical implementations will then be discussed in the last section.

\begin{figure}[t]
\centering \includegraphics[width=0.95\columnwidth]{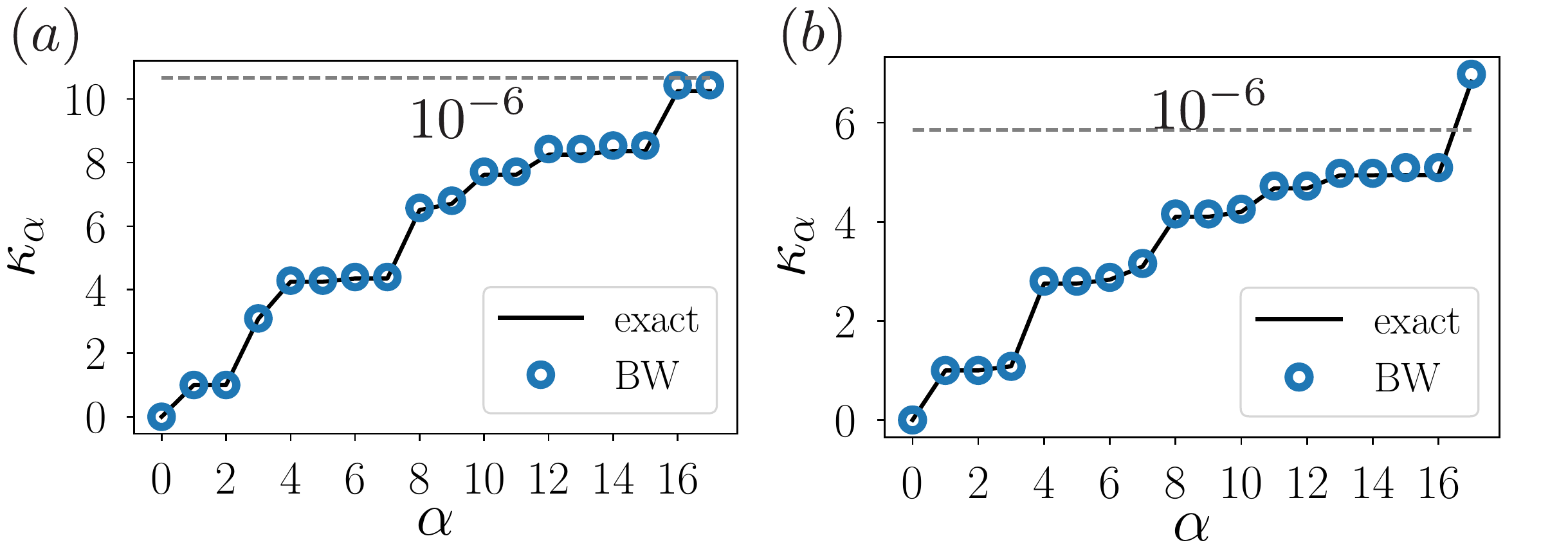}
\caption{{\it Entanglement Spectra of Heisenberg spin$-1/2$ chains.} We compare the BW prediction to the exact result for an OBC partition of $L'=24$ sites (we use $L=48$ for the exact calculation), with anisotropies $\Delta=-0.5$ (a) and $\Delta=0.9$ (b), finding excellent agreement even for moderate system sizes. Here, $\alpha_0=1$.
\label{fig:Heisenberg} }
\end{figure}

\paragraph*{Entanglement Hamiltonians of Heisenberg-type models. - } As a first case sample, we discuss the EH for spin-$1/2$ and spin-1 XXZ spin chains. For the $s=1/2$ case, the low-energy physics of $H_{\textrm{XXZ}}$ for $-1<\Delta\leq1$ is well described by a conformal field theory (CFT) with central charge $c=1$ (compactified boson)~\cite{fradkinbook}. In Fig.~\ref{fig:Heisenberg}a-b, we compare the exact spectrum (black line) obtained using density-matrix-renormalization-group (DMRG) simulations~\cite{schollwock2011density,White1992}, with the spectroscopy obtained using Eq.~\eqref{BWL}. On purpose, we show results with modest partition sizes of $L'=24$ sites, which are instrumental in view of the implementations discussed in the last section (for system sizes of order of 100 lattice sites, the agreement improves significantly). Down until eigenvalues $\tilde\lambda$ of order $10^{-6}$, the results of the universal ratios are almost undistinguishable, with errors at most at the $1\%$ level, despite the relative small sizes of the bipartitions. Similar results are obtained throughout the conformal phase, and even within the gapped, antiferromagnetic phase at $\Delta>1$ in the parameter regime where the ratio between correlation length and lattice spacing is of order 10.

For the $s=1$ case, the dynamics for $0<\Delta\lesssim1.2$ is captured by an $O(3)$ non-linear sigma model with topological angle $\theta=0$\cite{haldane1983nonlinear,fradkinbook}. The low-energy theory is Lorentz invariant, and the spectrum is gapped. The ground state displays symmetry-protected topological order - the so-called Haldane phase -, which is strikingly signalled by an (at least two-fold) degenerate entanglement spectrum~\cite{Pollmann:2010aa}. In Fig.~\ref{fig:recipe}e, we compare the numerically exact ES obtained via DMRG simulations, and the BW spectrum, also obtained via DMRG by targeting up to 10 states in each magnetisation sector with $|S^z| \leq 10$ for a bipartition of length $L'=40$. We show results obtained for a bipartition in a periodic system, using the corresponding BW adaption obtained in Ref.~\cite{Cardy:2016aa} for conformal field theories (the Haldane phase can be though of as a perturbed double sine-Gordon model~\cite{schulz1986phase}). In the left panel, we show how the characteristic degeneracies of the entanglement spectrum deep in the topological phase (here, we use $\alpha_0=4$ for clarity): for a sufficiently large systems, these degeneracies are perfectly captured, with deviations of the same order of the DMRG truncation errors we employed ($10^{-6}$). Moreover, close to criticality (right panel), the spectrum is also extremely well captured, and for bipartitions as small as $L'=8$, all eigenvalues with $\tilde\lambda<10^{-3}$ are well captured.

\begin{figure}
\centering \includegraphics[width=0.95\columnwidth]{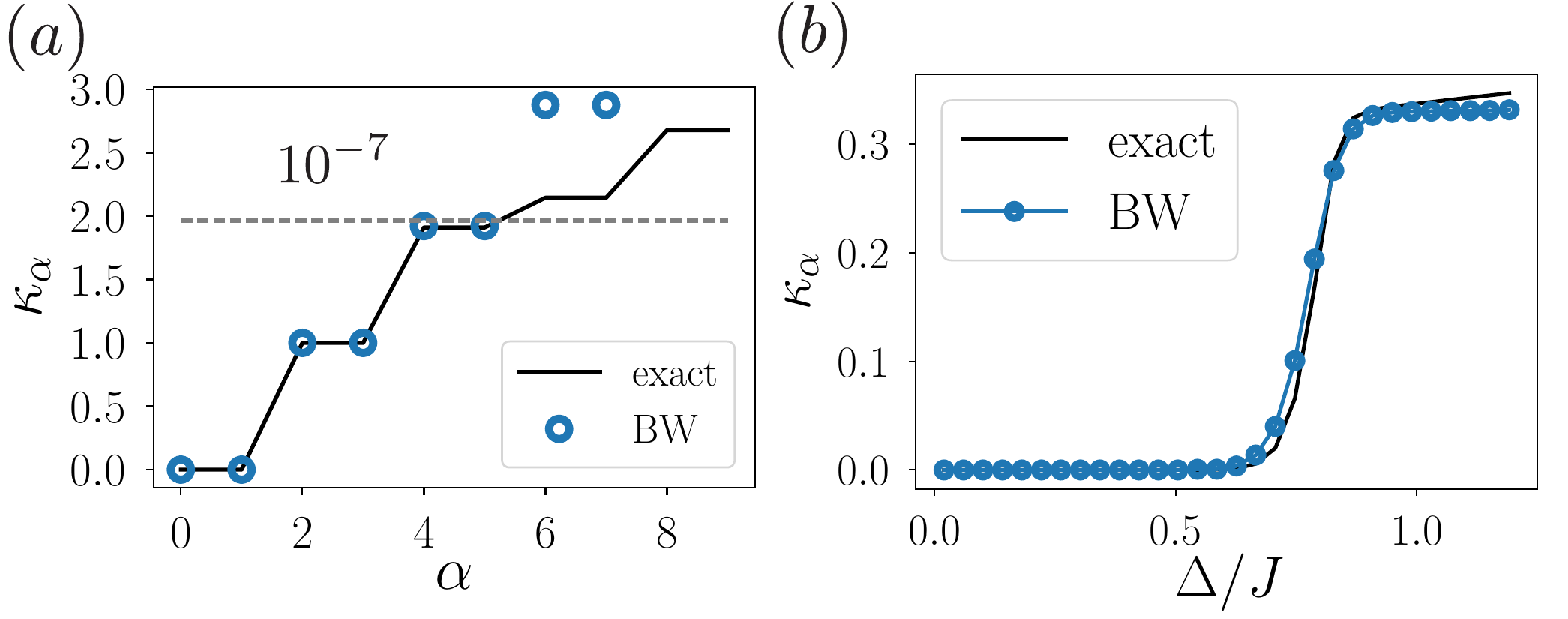}
\caption{{\it Entanglement Spectra of dipolar Ising chains}.
We consider an OBC partition of $L'=20$ sites. In panel (a), we show the entanglement spectrum for $\Delta=0.42J$, close to the critical point ($\alpha_0=2$). In panel (b), we represent the entanglement gap (aka Schmidt gap), renormalised with respect to the second excited state, as a function of the transverse field $\Delta$. The transition to the antiferromagnetic phase is shown by the closure of the entanglement gap.
\label{fig:ising} }
\end{figure}

\paragraph*{Entanglement Hamiltonians of Ising-type models. - } We now turn to models with discrete global $\mathbb{Z}_2$ symmetry, described by Ising-type models:
\begin{equation}\label{HIsing}
H_{IS} = J \sum_{n<p} \frac{1}{|n-p|^\eta} \sigma_n^x \sigma_p^x+\Delta\sum_n \sigma_n^z, 
\end{equation}
with antiferromagnetic interactions $J>0$ and $\eta>0$. At large $\Delta\gg J $, the ground state is a paramagnet with all spins pointing along the z direction, while at small $\Delta\ll J$, the system enters an antiferromagnetic phase. The system undergoes a phase transition between those, that for sufficiently large $\eta$ is described by a $c=1/2$ CFT (real fermion)~\cite{Koffel2012}. This transition is expected to have clear signatures in the low-lying ES: in particular, the entanglement gap should be open in the paramagnetic phase, and closed in the anti-ferromagnetic, symmetry broken one.

In Fig.~\ref{fig:ising}a, we compare the BW and exact ES in the anti-ferromagnetic phase for the case of dipolar interactions~\cite{Blatt2012,saffman2010quantum}, $\eta=3$, which are effectively short-ranged in 1D. 
Here, the BW Hamiltonian is given by $\tilde {\mathcal{H}}_{A,IS}  = J \sum_{n<p} \frac{n+p-1}{2|n-p|^\eta} \sigma_n^x \sigma_p^x+\Delta\sum_n  (n-\frac{1}{2} )\sigma_n^z$.
Degeneracies are clear and the agreement is excellent until the $10^{-7}$ level. In Fig.~\ref{fig:ising}b, we plot a scan of the entanglement gap as a function of $\Delta/J$: again, the agreement is very good even in the vicinity of the transition point, despite the moderate system sizes used here. In particular, both methods locate the finite-size transition point (corresponding to $\kappa_{1;2} \approx 0.15$). Remarkably, similar results are obtained for $\eta=1.5$, which displays an intermediate behaviour between long- and short-range interactions, as discussed in supplementary material (SM).

\paragraph*{Entanglement Hamiltonians for free fermions in one dimension.}

In this section, we consider free fermions in a one dimensional lattice are described by the Hamiltonian
\begin{equation}
H_f = - t \sum_{n=1}^{L} (c^\dagger_nc_{n+1} + \textrm{h.c.})
\end{equation}
where $c_{n}$ are fermionic annihilation operators at the site $n$. 
This model is ideal to benchmark our strategy as we can compute the exact (thermodynamic limit) ES based on the knowledge of the ground state correlation functions~\cite{Peschel2004,cheong2004many}. Moreover, the deviations from the linear dispersion relation characteristic of gapless (Lorentz invariant) fermions in the continuum can be conveniently tuned using the filling fraction, $\nu= N/L$, where $N$ is the total number of particles in the system. The latter element is particularly important here, as it provides a quantitative guide to address the role of such effects and, thus, illustrates the regimes of applicability of our technique.

\begin{figure}[t]
\centering \includegraphics[width=0.95\columnwidth]{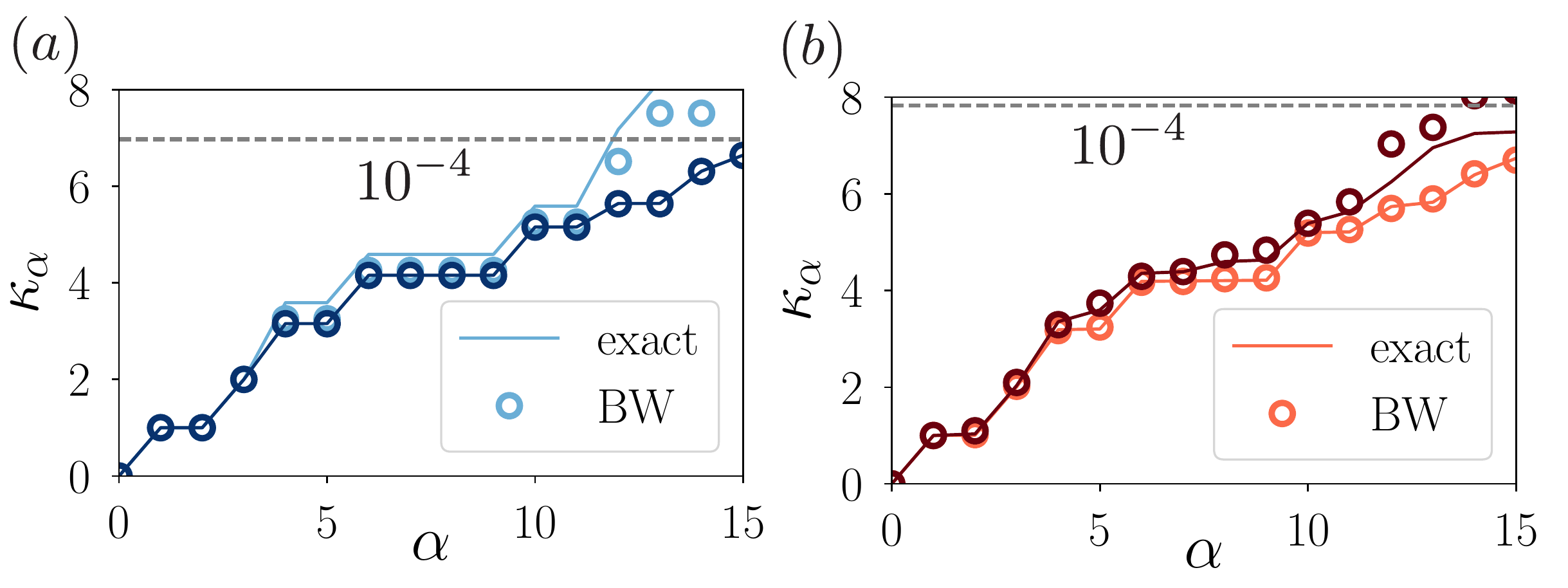}
\caption{{\it Entanglement Spectra of free fermions in 1D (a-b)}
For free fermions models, we study the effect of finite partition sizes in (a) with two PBC partitions of $L'=4$ (light blue) and $L'=32$ (dark blue), with density $\nu=1/2$.
In panel (b), we show how the BW prediction becomes less accurate at lower fillings. The light (dark) red line corresponds to $\nu=1/4$ ($\nu=1/32$). In both panels, $\alpha_0=1$.
\label{fig:1dfermions} }
\end{figure}

In Fig.~\ref{fig:1dfermions}a, we compare results obtained using the ES from the exact ground state in the infinite size limit $L\to\infty$, and finite size results using the BW theorem. Remarkably, even for system sizes as small as $L'=4$ (light blue), the first 4 eigenvalues are almost exactly matching. Large deviations take place relatively quickly after that. Going to $L'=32$, the errors become of order $0.1\%$ until $\tilde\lambda\sim 10^{-5}$. In Fig.~\ref{fig:1dfermions}b, we use $L'=32$, and compare different filling fractions, $\nu=1/4$ and $\nu=1/32$. Despite the overall good agreement, we note that the very dilute case shows deviations of order of $10\%$ already for relatively large eigenvalues, $\alpha\sim 5$. We attribute this discrepancy to the fact that, in this parameter regime, deviations from Lorentz invariance are more severe, as expected.

\begin{figure}
\centering \includegraphics[width=0.95\columnwidth]{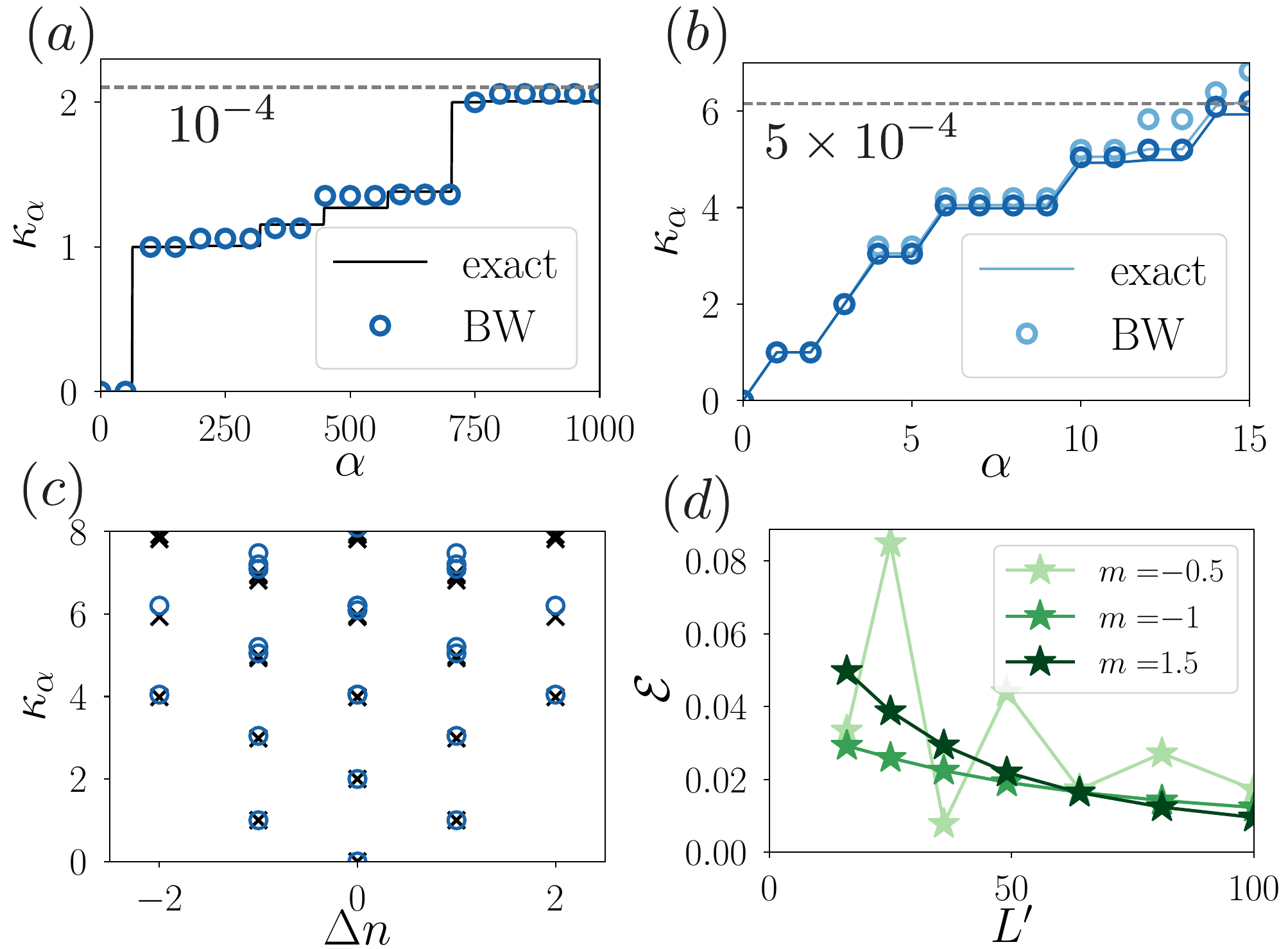}
\caption{{\it Entanglement Spectra in 2D hopping models (a) and topological insulators (b-c-d)}
(a) Free fermions in a hopping models with a PBC partition of $L'=36$ sites and density $\nu=1/2$ $\alpha_0=64$. The first dimensionless ratios obtained using the BW theorem agree with the exact result.
(b) ES for the massive Dirac model and in the topological phase (with $m=-1$), at unit filling, with square OBC partitions of $L'=16$ (light blue) and $L'=100$ (dark blue). Here, $\alpha_0=1$.
(c) Same as panel (b) for $L'=100$ where we represent the dimensionless ratio versus the quantum number, revealing at low energies the edge state spectrum.
(d) Errors of the BW theorem for the first 10 eigenvalues for different values of $m$ and as a function of partition size $L'$.
\label{fig:2d} }
\end{figure}

\paragraph*{Entanglement Hamiltonians for fermionic systems in two-dimensions. - } Finally, we analyze the accuracy of the BW scenarios for 2D systems. In two-dimensional lattices, there is a need of adapting the Bisognano-Wichmann theorem to finite geometries in two directions (in the continuum and in presence of conformal symmetry, the EH following BW can be obtained as in the 1D case~\cite{Casini:2011aa}). For this purpose, we employ a conformal mapping to derive the effective couplings of the EH for a square subregion (see SM): since the form of the EH stems from the properties of the light-cone coordinates, this choice is supposed to work well in our scenario. We remark that, for sufficiently large systems, it is still possible to use the original BW formulation, so the latter approximation should be mostly understood as an additional tool to further reduce experimental resources. We start from free, spinless fermions, focusing on small partitions of size $6\times 6$ accessible within the implementation discussed below. The corresponding results, shown in Fig.~\ref{fig:2d}a, illustrate how the first thousand eigenvalues are all within 10\% of the exact result. The error is of order of 1\% for the first 100 eigenvalues.

As a second example, we consider a 2D Dirac model~\cite{Qi2008}, defined as
\begin{eqnarray}
H_{MD}&=&\sum_\mathbf{n} \mathbf{c}^\dagger_{\mathbf{n}}\frac{\sigma_z-i\sigma_x}{2} \mathbf{c}_{\mathbf{n}+\mathbf{x}}  +\mathbf{c}^\dagger_{\mathbf{n}} \frac{\sigma_z-i\sigma_y}{2} \mathbf{c}_{\mathbf{n}+\mathbf{y}}+\mathrm{h.c.} \nonumber\\
			&+&m\sum_\mathbf{n} \mathbf{c}^\dagger_\mathbf{n} \sigma_z \mathbf{c}_\mathrm{\mathbf{n}}, 
\end{eqnarray}
where, $\mathbf{n}=(\mathbf{n}_x,\mathbf{n}_y)$ denotes a 2D index, the $\mathbf{c}_\mathbf{n}$ operators are spinfull fermions and $\sigma_{x,y,z}$ are the $2\times2$ Pauli matrices. In the following, we consider $-2<m<0$ for which the lowest band has Chern number $\mathcal{C}=-1$. In Fig.~\ref{fig:2d}b, we show that the low lying eigenvalues of the ES include degeneracies, associated with the edge state dispersion relation, that are perfectly resolved, in particular for a large (square) partition size $L'=100$. 
Note here that the BW prediction is directly related with the Li-Haldane conjecture~\cite{Li2008,Swingle2012}: the ES reveals the edge excitations of the model Hamiltonian, while containing information about the bulk.
In Fig.~\ref{fig:2d}c, we plot the comparison between ES as a function of the number of particles in the system ($\Delta n =0$ corresponds to unit filling): according to the bulk-edge correspondence, the spectrum is characterised by a linear dispersion relation, related to the spectrum of the gapless edge modes. The BW result is extremely accurate in reproducing quantitatively this feature, and further, it reproduces well also low-lying excited states in each $\Delta n$ sector. 

In this example, the accuracy of the BW theorem depends on the linearity of the edge state dispersion. This is illustrated in Fig.~\ref{fig:2d}d where we show the relative error $\mathcal{E}$ of the BW prediction, averaged over the first lowest ten ES values, and as a function of the number of sites $L'$ in the partition $A$. For $m=-1$, the dispersion relation of the edge states is approximately linear and the error $\mathcal{E}$ remains very small, even for very small partitions sizes. For $m=-1.5,0.5$ the nonlinear character of the edge state dispersion leads to slightly larger errors.

\begin{figure}[t]
\centering \includegraphics[width=0.45\textwidth]{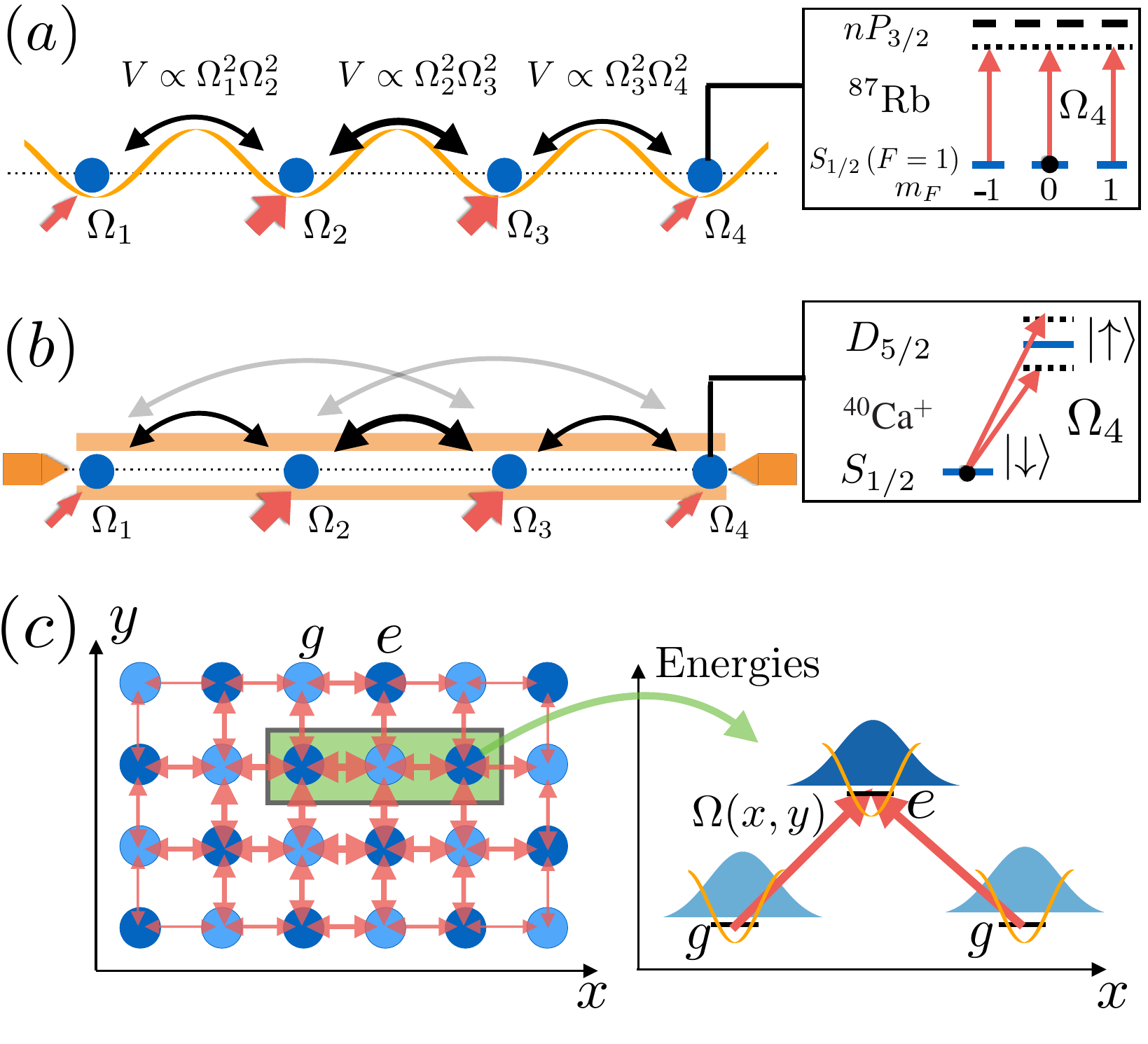}
\caption{{\it Implementations of entanglement Hamiltonians.}
(a) Realisation of the entanglement Hamiltonian of the Spin $1$ Heisenberg chain with Rydberg dressing. Spin states are encoded in a hyperfine ground state manifold (here $F=1$ of Rubidium atoms) and inhomogeneous interactions are obtained by off-resonant, spatially-dependent, laser excitations to a Rydberg state~\cite{VanBijnen2015}.
(b) Implementation of the long-range Ising models with trapped ions in a Paul trap. Inhomogeneous interactions are obtained by coupling electronic levels using spatial dependent laser couplings to phonon modes. 
(c) Realisation of free fermions Entanglement Hamiltonians via laser-assisted tunnelling in optical lattices. 
\label{fig:imp} }
\end{figure}

\section*{Implementation of Entanglement Hamiltonians and spectroscopy}

From the perspective of quantum engineering of entanglement Hamiltonians, the BW theorem guarantees that no exotic interactions are present in the EH $\tilde{\cal H}_A$, as the only difference with respect to $H$ are inhomogeneous couplings. This implies that, if one is able to engineer the system Hamiltonian with local control over couplings, also the corresponding BW EH can be realised. In particular, the AMO Quantum Simulation toolbox~\cite{Bloch2012,Blatt2012}, based on the trapping of atoms or ions and light-assisted interactions, provides all the necessary ingredients to implement the EH in state-of-the-art experimental setups, thus allowing direct measurement of the corresponding ES.

We show in Fig.~\ref{fig:imp} three illustrative examples of implementations of the BW EH:  the spin-$1$ XXZ model with Rydberg atoms [panel (a)], the long range Ising model with trapped ions [panel (b)] and a model of free fermions with ultracold atoms [panel (c)].
Note that spatially dependent interactions can also be realised using optical tweezers arrays with non constant atomic separations (enabling for instance the realization of the EH of the transverse Ising model with Rydberg atoms).
In all three cases, our implementations are based on the existing toolbox to realise spins and fermions models using light-assisted interactions~\cite{VanBijnen2015,Glaetzle2015,Porras2004,Kim2011,Jaksch2003,Gerbier2010}. Inhomogeneous couplings following the prescription of the BW theorem are then realised based on spatially dependent laser intensities. Additional details are presented in the SM.
This approach can be naturally adapted to implement the EH within others quantum simulation platforms, such as polar molecules~\cite{Micheli2006}, magnetic atoms~\cite{Lahaye2009}, and solid-state setups with NV centers~\cite{Cai2013} or with superconducting quantum circuits~\cite{Houck2012}.

Once the EH has beed engineered, the corresponding entanglement spectrum ${\tilde\epsilon_\alpha}$ can be measured using well established techniques based on many-body spectroscopy.  
An option, realised in a trapped ion setup~\cite{Senko2014} consists in preparing the ground state, or a excited state of the EH via adiabatic state preparation and monitoring the response of an observable $\langle O(t)\rangle$ to a weak perturbation $h(t)=h\sin(\nu t)$, where $h$ is a local operator. 
The resolution $\Delta\nu$ of the measurement is limited by the time of duration $t_\mathrm{obs}$ of the spectroscopy and the coherence time of the system $t_\mathrm{coh}$.

We applied this method to simulate the spectroscopy of the spin-$1$ XXZ model, for the parameters presented in Fig.~\ref{fig:recipe}e. The results are presented in panel (f) where we represent the root mean square (over time) of an observable $\langle S_1^\gamma(t)\rangle$ subject to a perturbation $h=aS_1^\gamma$ ($\gamma=x,y$), for as a function of the probe frequency $\nu$. 
Note that in this example, the fourth excited state is not coupled to the ground state by $h(t)$. This implies that the spectroscopy has also to be performed from the first excited state.
In the SM, we present another method to realise the ES spectroscopy based on the fast preparation of a superposition state of the low lying EH eigenstates, followed by Fourier analysis.

The spectroscopic method requires the initialisation of the ground state of EH - the entanglement ground state. Its adiabatic state preparation is equivalent to conventional ground state problems in synthetic quantum systems. The main difference is that here the time scale for state preparation is set by the smallest coupling in the system $J$. Depending on the system size and geometry of the bipartition, this is a factor of $L'$ smaller than the largest coupling available. 
We note that this does not depend at all on the dimensionality of the system, and that PBC of the bipartition further help in decreasing the ratio between largest and smallest coupling. Moreover, since we are only interested in spectral properties and not in ground state correlations, alternative to spectroscopy on GS exists. Finally, we remark that the procedure we employ is robust against finite temperature effects and the presence of noise during the spectroscopy.  These effects lead, as in conventional spectroscopy, to a broadening of the eigenvalue peaks without affecting their positions.
Note that the  $1/L'$ effect mentioned above is also relevant regarding decoherence rates, which should be compared to the smallest coupling $J$ to assess the important of decoherence mechanisms. This is illustrated in the case of 1D Ising model with dephasing in the Supplementary material.

\section*{Outlook}

Our proposal can be realised in state of the art atomic, optical and solid state experimental setups and immediately extended to investigate entanglement features beyond the ES: this includes the behaviour of correlations in the entanglement ground state(s) at quantum critical points, and other key quantities such as relative entropies~\cite{Casini:2014aa}. At the theoretical level, our approach immediately motivates new connections, based on experimental feasibility, to understand the structure of entanglement spectra in many body systems. This include a deeper understanding of bulk effects on the entanglement spectrum of both topological and critical theories, the role of different geometries in determining the functional form of the entanglement Hamiltonian within axiomatic field theory, and the ability of the entanglement spectrum to detect~\cite{Chiara:2012aa} or miss~\cite{chandran2014universal} quantum phase transitions. Finally, by creating a new bridge between axiomatic quantum field theory from one side, and synthetic quantum systems from the other, the strategy we put forward immediately motivates new field theoretical approaches to obtain the entanglement Hamiltonian of quantum field theories beyond ground state physics, such as quantum quenches and thermal states, which have very recently drawn attention in low-dimensional systems~\cite{Cardy:2016aa}.

\paragraph*{Acknowledgments. - } We thank V. Alba, P. Calabrese, L. Chomaz, R. Fazio, C. Roos, E. Tonni and R. van Bijnen for useful discussions. MD thanks M. Falconi for useful discussions and clarifications on Ref.~\onlinecite{guido2011modular}. 
Some of the DMRG simulations were performed using the ITensor library (http://itensor.org) and simulations of the entanglement spectroscopy were carried out with QuTiP~\cite{Johansson20131234}.
Work in Innsbruck was supported in part by the ERC Synergy Grant UQUAM, SIQS, and the SFB FoQuS (FWF Project No. F4016-N23).

\appendix

\section{Entanglement Hamiltonians  in two dimensions}

In the case of two-dimensional models, for the sake of experimental implementation, one has to deal with an additional finite boundary effect not present in the one-dimensional case (where the boundary is only one site and the only correction stems from the finite size of the partition). In particular, we have to consider the finite size of the boundary, and the fact that, on the lattice, the partition cannot have an exact spherical shape (in the continuum, the EH for such case is known in the case of conformal field theories, see e.g. Ref.~\cite{Casini:2011aa}). 

For two-dimensional models and for partitions placed at the center of the system (c.f. for example Fig.~5 in the main text), we cast the BW theorem in the form
\begin{eqnarray}
H&=&\sum_{\mathbf{n}} H_{\mathbf{n},\mathbf{x}} + H_{\mathbf{n},\mathbf{y}}\\
 \tilde {\mathcal{H}}_A&=&\sum_\mathbf{n=1}^{L'^2} d(x_\mathbf{n}+1/2,y_\mathbf{n}) H_{\mathbf{n},\mathbf{x}} + d(x_\mathbf{n},y_\mathbf{n}+1/2) H_{\mathbf{n},\mathbf{y}} \nonumber,
\end{eqnarray}
with $H_{\mathbf{n},\mathbf{x}}$ ($H_{\mathbf{n},\mathbf{y}}$) represent the interaction terms between sites $\mathbf{n}$ and $\mathbf{n}+\mathbf{x}$ ($\mathbf{n}+\mathbf{y}$). Here, $x_\mathbf{n},y_\mathbf{n}$ are the lattice coordinates in units of the lattice spacing, defined with respect to the center of the partition. The inhomogeneity is written as $d(x,y)=(R^2-r^2(x,y))/(2R)$, with $R=L'/\sqrt \pi$ and $r(x,y)$ is the conformal distance from the center of partition to the lattice point $(x,y)$, calculated according to 
\begin{eqnarray}
r(x,y) &=& \sqrt{x'^2+y'^2} \\
x' &=& \mathrm{Re} (\frac{1-i}{\sqrt{2}} w)R\\
y' &=& \mathrm{Im} (\frac{1-i}{\sqrt{2}} w)R\\
w &=& c_n(K\frac{1+i}{2R} (x+iy)-K,\frac{1}{2}), 
\end{eqnarray}
with $K\approx 1.854$  and $c_n$ a Jacobi elliptic function. This choice is qualitatively justified by the fact that the conformal distance well approximates the linear relation between space and time for excitations generated by the bipartition (even in the absence of conformal symmetry). We remark that the full BW procedure can still be applied if one consider very large partitions, with $L_x\gg L_y$, which essentially mimic the original half-plane geometry.

\section{Details about the Implementation of Entanglement Hamiltonian}

Here, we provide additional details the AMO implementations of Entanglement Hamiltonians, as depicted in Fig.~6 in the main text.

\subsubsection{Rydberg atoms}
In the first panel (a), we depict the implementation of the Entanglement Hamiltonian of a spin$-1$ XXZ model with Rydberg-dressed atoms, based on the ideas developed in ~Ref.~\cite{VanBijnen2015}.
We consider a chain of atoms $n=1,..,L'$,  in a Mott insulating phase and encode the states of the spin one model in a hyperfine ground state manifold of, for instance, Rubidium atoms: $\ket{-1,0,1}\equiv \ket{5S_{1/2}F=1,m_F=-1,0,1}$, with a magnetic field aligned in the direction of the atoms defining the quantization axis.
Interactions between spins are realized by Rydberg dressing~\cite{Henkel2010,Pupillo2010}, which consists in exciting off-resonantly the hyperfine states to Rydberg states, which here belong to a $P_{3/2}$ fine-structure manifold (with very large first quantum number). 
In fourth-order perturbation theory in the small parameters $\Omega_n/\Delta\ll1$, where $\Omega_n$ are spatially dependent Rabi frequencies and $\Delta$ is the laser detuning, one can obtain the Hamiltonian of the form~\cite{VanBijnen2015}
\begin{equation}
H= \sum_{n<p} \left[J_{n,p} (S^x_nS^x_{p} + S^y_nS^y_{p})+\Delta_{n,p} S^z_nS^z_{p})\right], 
\end{equation}
with $J_{n,p}=\Omega_n^2 \Omega_p^2 f(r_{np})$, and $\Delta_{n,p}=\Omega_n^2 \Omega_p^2 g(r_{np})$. The functions $f$ and $g$ depend on the polarization and detuning of the laser beams and on the Rydberg state manifold (atom, first quantum number), and can be in particular parametrized to obtain in good approximation nearest-neighbor interactions: $f(r_{np})=\delta_{p,n+1} f$, $g(r_{np})=\delta_{p,n+1} g$, leading to $J_{n,p}= \delta_{p,n+1} f \Omega_n^2 \Omega_{n+1}^2 $, $\Delta_{n,p} =\delta_{p,n+1}g \Omega_n^2 \Omega_{n+1}^2 $
Finally, these couplings can be made inhomogeneous in order to realize the EH following the prescription of the BW theorem by using spatially dependent Rabi Frequencies $\Omega_n$.

\subsubsection{Trapped ions}
As shown in Fig.~6(b), the same approach applies to realize Entanglement Hamiltonians of long-range Ising models with trapped ions.
For each ion $n=1,..,L'$, the two levels representing a pseudo spin $1/2$ particle are encoded in two long-lived electronic states, with transition frequency $\omega_0$ .
The spins are coupled to the transverse modes $m=1,..,2L'$ of the ion chain with frequencies $\nu_m$ using bichromatic laser beams with frequencies $\omega=\omega_0\pm \Delta$, Rabi frequencies $\Omega_n$ and in the Lamb-Dicke regime with the Lamb-Dicke factors $\eta_{n,m} \ll1$~\cite{Blatt2012}.
In the weak-coupling limit, $\eta_{n,m}\Omega_n \ll \Delta$. we obtain in the second-order perturbation theory a coupling between spins in the form of an Ising interaction $H=J_{np}\sigma_n^x\sigma_p^x$~\cite{Porras2004,Kim2011} with
\begin{equation}
J_{np}=- \sum_{m} \frac{\Omega_n \Omega_p \eta_{n,m} \eta_{p,m}\nu_m}{\Delta^2-\nu_m^2}.
\end{equation}
For homogenous laser beams, $\Omega_n=\Omega$, the interaction matrix can be approximated by a power law $J_{np}=J/|n-p|^\eta$, where $\eta$ can be tuned between $0$ and $3$~\cite{Blatt2012}, allowing to realize the Hamiltonian of the dipolar Ising model.  The corresponding EH can be then implemented using spatially dependent Rabi frequencies [c.f Fig.~6(b)] for the Ising terms. Finally, a magnetic field gradient can be used to implement the longitudinal field.

\subsubsection{Ultracold fermions}
As a last illustration, we show how to engineer the Entanglement Hamiltonian of free fermion models, with ultracold atoms hopping in a optical lattice via laser-assisted tunnelling~\cite{Jaksch2003,Gerbier2010}.
In Fig.~6(c), we represent the setup we have in mind with fermionic atoms with can be either in a ground state level $\ket{g}$ or a long-lived clock state $\ket{e}$.
The atoms can move in a two-dimensional square lattice of lattice period $a$, where they occupy the state $\ket{g}$ on the sites $\mathbf{n}=(x_\mathbf{n},y_\mathbf{n})$ with $x_\mathbf{n}+y_\mathbf{n}$ even, and the states $\ket{e}$ on the other sites [cf Fig.~6(c)]. This can be achieved with laser beams forming a checkerboard optical lattice and where atoms in the state $\ket{g}$ (respectively $\ket{e}$) are trapped in a laser intensity minimum (maximum).

Hoppings between sites accommodating different states are obtained by laser-assisted tunneling, which consists in coupling resonantly via a laser-beam the two levels $\ket{e}$ and $\ket{g}$. Considering only nearest neighbor interactions, we obtain the free-fermion Hamiltonian~\cite{Jaksch2003,Gerbier2010}
\begin{equation}
H = \sum_\mathbf{n} t_{\mathbf{n},\mathbf{x}} c^\dagger_\mathbf{n} c_{\mathbf{n}+\mathbf{x}}+t_{\mathbf{n},\mathbf{y}} c^\dagger_\mathbf{n} c_{\mathbf{n}+\mathbf{y}}+\mathrm{h.c}
\end{equation}
with $t_{\mathbf{n},\mathbf{m}}=\int \Omega(\mathbf{r}) \omega_\mathbf{n}(\mathbf{r}) \omega_{\mathbf{n}+\mathbf{m}}(\mathbf{r}) d\mathbf{r}$, where $\omega_\mathbf{n}$ denotes the Wannier function at site $\mathbf{n}$. As in the previous examples, the spatial control of the Rabi frequencies $\Omega(\mathbf{r})$, which we consider here real for simplicity, allows to implement the EH according to the BW theorem.

\section{Ising models with power-law interactions}

In the main text, we represent the ES of the quantum Ising model with dipolar interactions. Here we present the case of the power law exponent is $\eta=1.5$, which can also be realized in trapped ions setups~\cite{Blatt2012}. We remark that, for $\eta>1$, interactions are effectively short-ranged, while in the opposite regime, we do not expect the BW theorem to hold due to true long-range correlations (which affect the dispersion relation of excitations in a drastic manner).

\begin{figure}[h]
\centering \includegraphics[width=0.95\columnwidth]{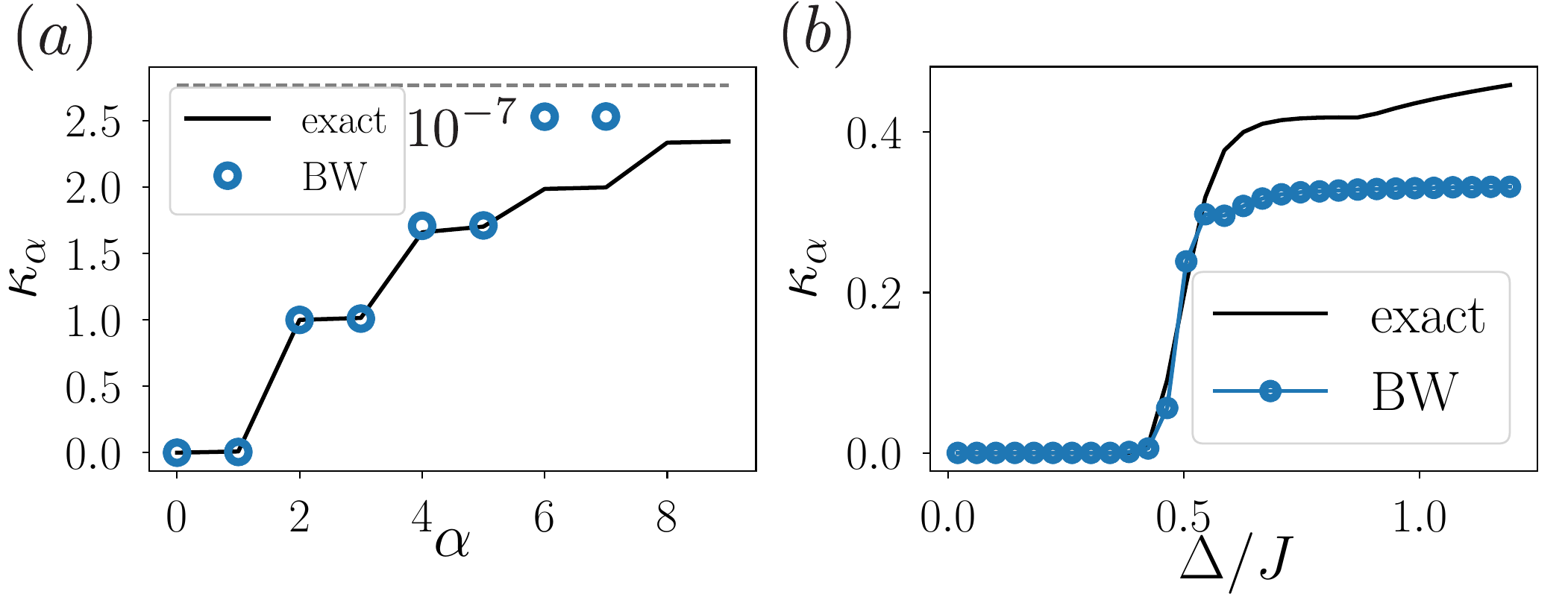}
\caption{{\it Entanglement Spectra of Ising chains with power-law exponent $\eta=1.5$}.
We consider an OBC partition of $L'=20$ sites. In panel (a), we show the entanglement spectrum for $\Delta=0.42J$, close to the critical point. In panel (b), we represent the entanglement gap (aka Schmidt gap), renormalized with respect to the second excited state, as a function of the transverse field $\Delta$. The transition to the antiferromagnetic phase is shown by the closure of the entanglement gap.\label{fig:ising15} }
\end{figure}

The BW prediction agrees well with the exact result, and in particular, we find excellent quantitative agreement in the antiferromagnetic phase and in the vicinity of the critical point - see Fig.~\ref{fig:ising15}. Deep in the paramagnetic phase, we observe sizeable deviations: we attribute the latter to two features of the model. First, finite size effects are expected to be considerably larger here with respect to the $\eta=3$ case, where the ES was well captured also in the paramagnetic phase. Second, power-law interactions, while keeping locality at an effective level, introduce also non-universal features, such as additional power-law corrections to correlation functions. The latter might not be well captured within BW, and can thus lead to quantitative deviations when comparing the spectra.

\section{Diabatic spectroscopy and decoherence effects}
\begin{figure}
\centering \includegraphics[width=0.95\columnwidth]{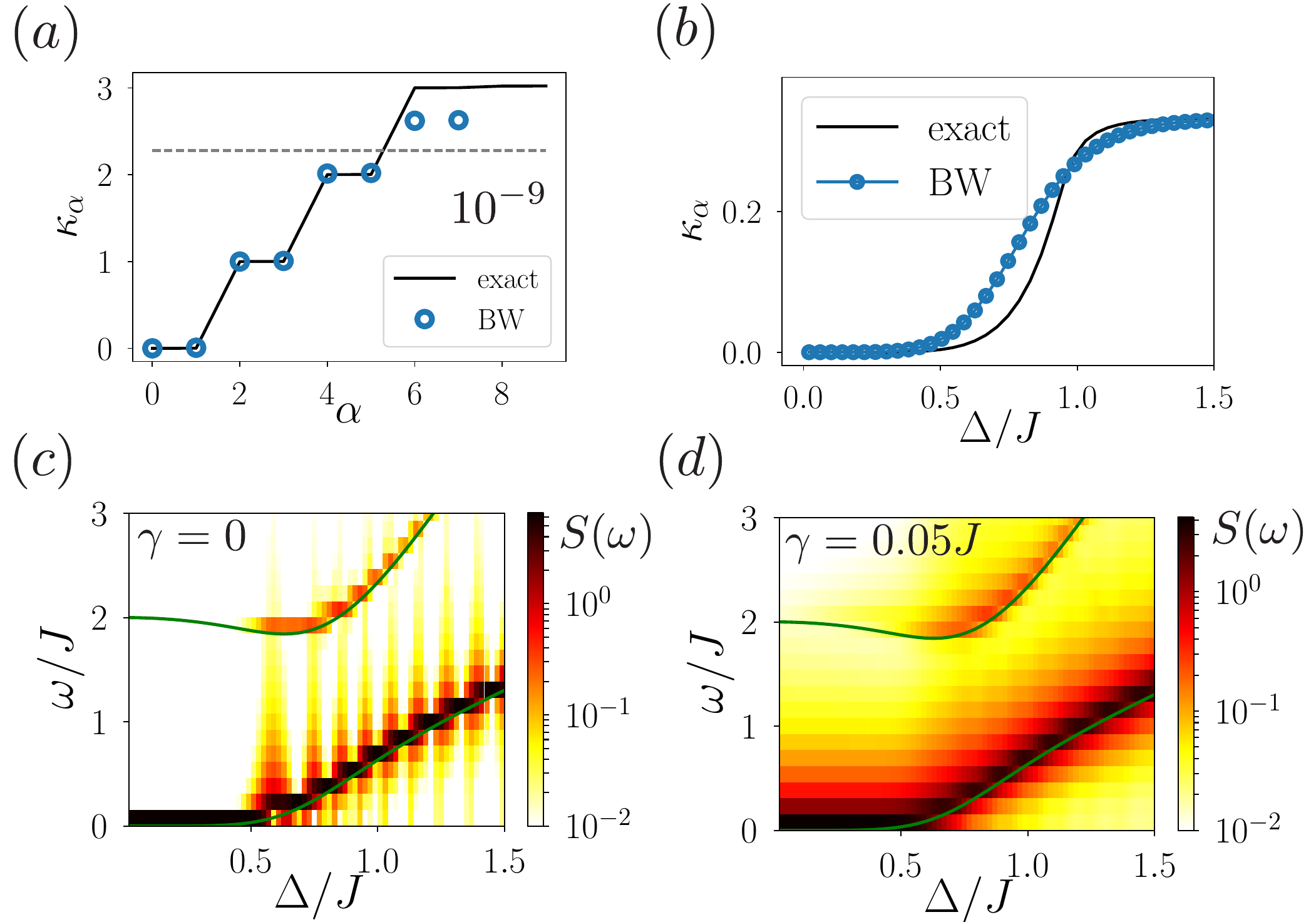}
\caption{{\it Entanglement Spectra of Ising chains with NN interactions and corresponding diabatic spectroscopy.}
We consider an OBC partition of $L'=6$ sites. In panel (a), we show the entanglement spectrum for $\Delta=0.42J$, close to the critical point. In panel (b), we represent the entanglement gap (aka Schmidt gap), renormalized with respect to the second excited state, as a function of the transverse field $\Delta$. 
In panels (c) and (d), we show the corresponding diabatic spectroscopy as a function of the transverse field $\Delta$. 
The color plot represents the normalized spectrum of the observable $\langle \sigma_1^z (t) \rangle$ and the solid lines correspond to the eigenvalues of the EH following the BW theorem.
We considered $c=10$, $J t_f=2$, $J t_\mathrm{obs}=40$ for dephasing rates $\gamma=0$ [panel (c)] and $\gamma=0.05J$ [panel (d)].
\label{fig:diabatic} }
\end{figure}

In this section, we present a method to realize the spectroscopy of the Entanglement Hamiltonian, which is an alternative to the linear response approach.
This method is referred as diabatic spectroscopy where, instead of preparing the ground state of the EH, we excite a superposition of low-lying excited states~\cite{Yoshimura2014}. The beating between the corresponding energies will reveal the ES.

The Hamiltonian we have in mind is written as
\begin{equation}
H(t)= \tilde {\mathcal{H}}_A +f(t)h +g(t) h', 
\end{equation}
with $f(t)=c\cos^2(\frac{\pi t}{2t_f})$, $g(t)=c\sin(\frac{\pi t}{t_f})$, $c\gg1$ and $t_f$ is the time of the preparation.
The Hamiltonian $\tilde {\mathcal{H}}_A$ is the Entanglement Hamiltonian we are interested in, $h$ is a Hamiltonian whose ground state can be prepared experimentally and $h'$ is an additional term which can be useful to connect eigenstates with different symmetries (see below).
In the following, we will consider the example of the Ising model with nearest neighbor interactions:
\begin{equation}
\tilde {\mathcal{H}}_A = J \sum_{n=1}^{L'-1} n \sigma_n^x \sigma_{n+1}^x + \Delta \sum_{n=1}^{L'} (n-\frac{1}{2}) \sigma_n^z, 
\end{equation}
and 
\begin{equation}
h= J \sum_n \sigma_n^z \quad h'= J \sum_n \sigma_n^x.
\end{equation}
Here, $h'$ breaks the parity symmetry in order to excite the first excited state of $\tilde {\mathcal{H}}_A$.

At time $t=0$, $H(t)\approx h$ and we initialize the system in the ground state of $h$, which is here $\ket{G}=\ket{\downarrow..\downarrow}$. We perform then a fast "diabatic" ramp following $H(t)$ with $Jt_f \sim 1$ resulting in a superposition of low-lowing excited state. The second part of the protocol consists in monitoring the dynamics of some observables $\langle O(t) \rangle$ during a time $t_\mathrm{obs}$. The ES is then visible in the corresponding spectrum $S(\omega)$.

In Fig.~\ref{fig:diabatic}, we show the results of the diabatic spectroscopy. In the absence of decoherence mechanisms, very clear peaks, with widths limited by the observation time $t_\mathrm{obs}$, emerge, revealing the Entanglement Spectrum. In presence of dephasing, the width becomes also limited by the corresponding rate $\gamma$ (see panel (d)). The ES is accessible by diabatic spectroscopy provided $\gamma \ll J$ (Note that $J$ is the smallest coupling required to create the ES and $J(L'-1)$ the largest.)

%

\end{document}